\documentclass[aps,pre,twocolumn,groupedaddress,showpacs,amssymb,amsmath]{revtex4-1}
\usepackage{graphicx}
\usepackage{epstopdf}
\usepackage{braket}
\usepackage{color}
\usepackage{amsmath}
\usepackage{amssymb}
\definecolor{dblue}{rgb}{0,0,0.75}
\definecolor{dred}{rgb}{0.6,0,0}
\definecolor{dgreen}{rgb}{0,0.5,0}

\begin{document}
	
\title{Topological phase transitions in Chern insulators within three-band models}
\author{P. Potasz}
\email{pawel.potasz@pwr.edu.pl}
\author{B. Jaworowski}
\author{M. Kupczy\'{n}ski}
\author{M. Brzezi\'{n}ska}
\author{P. Kaczmarkiewicz}
\author{A. W\'{o}js}

\affiliation{Department of Theoretical Physics, Faculty of Fundamental Problems of Technology, Wroc\l{}aw University of Science and Technology, 50-370 Wroc\l{}aw, Poland}
\date{\today}

\begin{abstract}
We investigate topological phase transitions in Chern insulators within three-band models, focusing on the empty band and lowest band populated by spinless fermions. We consider Lieb and kagome lattices and notice phase transitions driven by the hopping integral between nearest-neighbors, which leads to the change of the lowest band Chern number from $C=1$ to $C=-1$. In the single-particle picture, different phases are examined by investigating corresponding entanglement spectra and the evolution of the entanglement entropy. Entanglement spectra reveal the spectral flow characteristic of topologically nontrivial systems before and after phase transitions. For the lowest band with $1/3$ filling, Fractional Chern insulator (FCI) phases are identified by examining the ground state momenta, the spectral flow, and counting of the entanglement energy levels below the gap in the entanglement spectrum. A quasilinear dependence of the entanglement entropy $\alpha(n_A)$ term is observed for FCI phase, similarly to linear behavior expected for Laughlin phase in the fractional quantum Hall effect. At the topological phase transition, for both empty and partially filled lowest bands, the energy gap closure and a discontinuity in the entanglement entropy are observed. This coincides with the divergence of a standard deviation of the Berry curvature. We notice also a phase transition driven by the nearest-neighbor interaction in the Lieb lattice, where the many-body energy gap closes. The phase transitions are shown to be stable for an arbitrary system size, thus predicted to be present in the thermodynamic limit. While our calculations are performed for an interaction energy far exceeding the gap between the two lowest energy bands, we note that the higher band does not affect the phase transitions, however destabilizes FCI phases. 
\end{abstract}
\maketitle

\section{Introduction}
Two-dimensional lattice models with topologically nontrivial bands characterized by a topological invariant, Chern number, have recently attracted significant attention \cite{Sun, Neupert, Tang, ChernExperiment, ChernExperiment2}. Since Haldane proposal of a Chern insulator on a honeycomb lattice \cite{Haldane}, a number of multiband models with parameters controling the band topology has been proposed, including kagome \cite{flatkagome, Tang, Zoology}, Lieb \cite{We}, pyrochlore \cite{trescher2012flat}, dice \cite{wang2011dice} and other lattices \cite{flatstar, flatsqoct}. Band topology is determined by the competition between trivial and nontrivial gap opening terms \cite{Haldane}. For two-band models, a topological phase transition corresponds to a transition between a system with two bands with opposite Chern numbers and two bands with zero Chern numbers, as a sum of Chern numbers of all the bands must be zero. Multiband models allow for more possibilities of transitions between bands with a variety of Chern numbers. 

Flattening of Chern insulator bands enhances the role of particle interactions. An apperance of strongly correlated gapped phases is expected in the case of fractional filled topological bands, in analogy to the Fractional Quantum Hall effect (FQHE) on Landau levels \cite{tsui, laughlin, Jain}. Calculations for spinless fermions within the exact-diagonalization method \cite{Neupert, SunNature, Zoology, PRX, KourtisTriangular, YangDisorder, highchern1, highchern2, highchern3, highchern4} and density matrix renormalization group (DMRG) \cite{cincio2013characterizing, liu2013bulk, Johannes} approaches have proved the existence of Laughlin-like phases for $1/3$ filling factor and later on for an entire family of other fractions \cite{BeyondLaughlin,hierarchy}. These phases, commonly called Fractional Chern insulators, can be identified by looking at the quasi-degeneracy of the ground state \cite{wen, quasideg}, the presence of the spectral flow \cite{sflow1, LaughlinArgument}, many-body Chern number \cite{sflow1}, the momentum counting of the ground state and quasihole spectrum \cite{HaldaneCounting,BernevigCounting,RevBergholtz,RevParameswaran,RevNeupert}. 

Entanglement measures have been shown as a promising direction for identification of topological phases \cite{Zozulya1, Zozulya2, Entanglement,HaldaneCounting, BernevigCounting}. Entanglement between two parts of the system allows to extract the edge properties of the system and excitations directly from the bulk ground state. In the single-particle case, the signature of a nontrivial phase is the spectral flow in the orbital entanglement spectrum (ES) \cite{Peschel, Vish:inv, Fritz:ES, Bernevig:Disord, Alex:CM, PhysRevB.82.085106}. For correlated phases, particle entanglement spectrum reveals the exclusion statistics inherent in excitations of such states \cite{Entanglement, EntanglementExcitation, ParticleEntanglement, Zoology}. The number of the entanglement energy levels below the entanglement gap satisfies the counting principle for a given Laughlin-like state. 

Our goal in this work is to analyze phase transitions between topologically nontrivial phases in the three-band models within single-particle picture and for a partially filled band using entanglement measures. We focus on the phase transitions in Chern insulators on the Lieb and kagome lattices. We study a non-interacting case corresponding to empty bands and many-body effects for fractionally occupied bands with $1/3$ filling factor. We investigate bands evolution with a nearest-neighbor hopping parameter and notice transitions of the lowest band between two topologically nontrivial phases, with Chern numbers $C=1$ and $C=-1$. We study the entanglement spectra within both phases and the evolution of the entanglement entropy. For a partially filled band, we distinguish regions with FCI and other unidentified phases. Topological phase transitions are investigated by studying the low energy many-body spectra. Universality of the results is confirmed by considering different systems sizes and an effect of interaction with a higher band.  

\section{Three-band models of Chern insulators}
\begin{figure}
	\centerline{\includegraphics[width=0.5\textwidth]{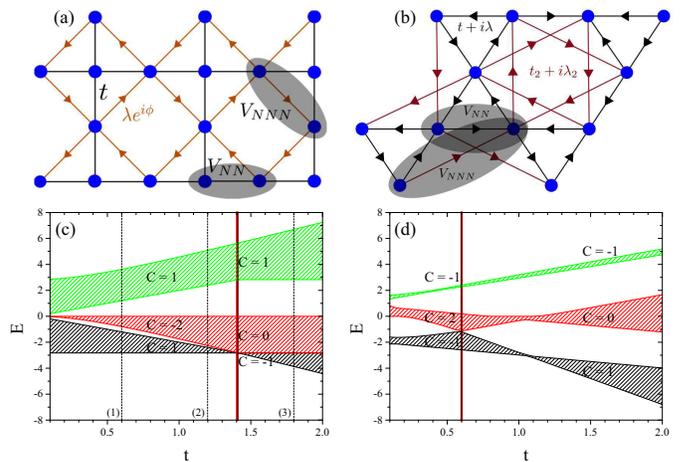}}
	\caption{(a) Lieb lattice with nearest-neighbor hopping $t$ and complex next-nearest neighbor hopping $\lambda e^{i\phi}$ marked with solid black line and brown lines with arrows, respectively. (b) Kagome lattice with complex nearest- (black lines) and next-nearest (brown lines) hopping integrals. (c) and (d) Evolution of the energy bands as a function of parameter $t$ for Lieb and kagome lattices, respectively. Topological phase transitions are indicated by the red vertical lines.} 
	\label{fig:Fig1}
\end{figure}

We consider two Chern insulator models with three atoms in a unit cell, which give rise to three energy bands after diagonalization the Hamiltonian. The first model is the Lieb lattice, a two-dimensional lattice with an additional site between each pair of the nearest-neighbors of a simple square lattice, as shown in Fig. \ref{fig:Fig1}(a) \cite{Weeks, Aldea, Zhao, Goldman, We}. In the tight-binding model, we include the nearest-neighbor hopping term $t$, the next-nearest neighbor hopping term $\lambda$ with an accumulated extra complex phase $\phi_{ij}=\pm \phi$ between sites $i$ and $j$ when hopping around a lattice site clockwise ($+$) and counterclockwise ($-$). Hamiltonian is written as
\begin{equation}
H=t\sum_{\braket{i,j}}c^{\dagger}_i c_j+\lambda\sum_{\braket{\braket{i,j}}} e^{i\phi_{ij}}c^{\dagger}_{i} c_{j},
\label{eq:Ham}
\end{equation}
where $c^{\dagger}_{i}$ ($c_{i}$) is creation (annihilation) operator acting on the site $i$. We set $\phi=\pi/4$ and $\lambda=1$ \cite{Neupert}.
 
The second model is the kagome lattice shown in Fig. \ref{fig:Fig1}(b), with complex hopping integrals between the nearest and next-nearest neighbors. Hamiltonian of the system reads
\begin{equation}
H=\sum_{\braket{i,j}}(-t+i\lambda_1w_{ij})c^{\dagger}_i c_j+\sum_{\braket{\braket{i,j}}}(t_2+i\lambda_2w_{ij})c^{\dagger}_{i} c_{j},
\label{eq:Ham}
\end{equation}
with parameters $\lambda_1=0.6$, $t_2=0.3$ and $\lambda_2=0.1$ chosen in a way that flattens the lowest band, and $w_{ij}=\pm 1$ when going clockwise ($+$) and counterclockwise ($-$) between sites $i$ and $j$.
We investigate topological properties of energy bands when a hopping integral $t$ between the nearest neighbors is turned on. The Lieb lattice at $t=0$ can be considered as a lattice consisting of a checkerboard lattice with complex hoppings (sites connected by the brown lines with arrows) completely decoupled from the lattice formed from individual sites in the center of the brown rhombs, see Fig. \ref{fig:Fig1}(a). 
We obtain two dispersive energy bands localized on a checkerboard lattice and a completely flat band at $E=0$ in the middle of them, which corresponds to isolated sites. All three bands touch each other at $E=0$. Turning on the hopping integral $t$ opens up the energy gaps between three bands. The evolution of band widths is schematically shown in Fig. \ref{fig:Fig1}(c). Calculating the Berry curvature for each energy band and integrating it over the entire Brillouin zone allows us to determine their Chern numbers. For small values of $t$, the lowest and the highest bands have Chern number $C = 1$, while the middle one has $C = -2 $. Here, we notice that infinitesimally small coupling of isolated sites to checkerboard lattice leads to topological phase transition. Isolated sites in the center of the brown rhombs corresponds to the atomic limit - a trivial insulator with $C=0$. Infinitesimally small coupling to a checkerboard lattice changes its topology to $C=-2$.  An increase of the value of $t$ leads to a stronger coupling of the isolated sites and a checkerboard lattice. At $t\sim 1.4$, we observe topological phase transition between the two lowest bands, manifested by the change of Chern numbers from $C=1$ to $C=-1$ (for the lowest band) and from $C=-2$ to $C=0$ (for the middle one). 
Similar phase transition is seen in the case of a kagome lattice, depicted in Fig. \ref{fig:Fig1}(d). For small values of $t$, the lowest and the highest bands have $C=-1$, and the middle one $C=2$. For $t\sim 0.55$, a phase transition occurs between two lower bands. The lowest band changes its Chern number to $C=1$ and the middle one to $C=0$. 

\begin{figure}
	\centerline{\includegraphics[width=\columnwidth]{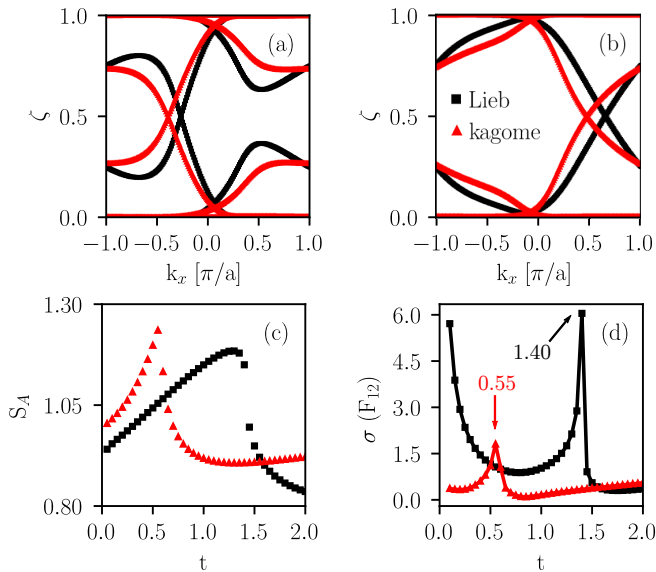}}
		\caption{Single-particle entanglement spectrum of the lowest band (a) before and (b) after phase transitions for Lieb lattice (black squares) and kagome lattice (red triangles). (c) Single-particle entanglement entropies as a function of hopping integral $t$. (d) Standard deviations of Berry curvature of the lowest band as a function of hopping integral $t$.}
	\label{fig:Fig2}
\end{figure}

Besides determining the Chern number of bands, we also verify nontrivial topological properties of the lowest band by investigating the entanglement spectra. We consider a system in the many-body ground state $\ket{\Psi_\mathrm{GS}}$, which is given by a Slater determinant of all the single-particle states from the lowest band. The density matrix of the composite system in this state is $\rho=\ket{\Psi_\mathrm{GS}}\bra{\Psi_\mathrm{GS}}$. We divide the system into two equal spatial parts, A and B in a torus geometry \cite{Alex:CM}. The reduced density matrix of the subregion A is defined as $\rho_A = \mathrm{Tr}_B \rho$. In the non-interacting case, the entanglement spectrum of a subsystem can be obtained from the correlation matrix \cite{Peschel}
		\begin{equation}
			C_{ij} = \braket{c_i^\dagger c_j}=\mathrm{Tr}_B(\rho_Ac_i^\dagger c_j),
		\end{equation}
where $\braket{}$ denotes the expectation value in the many-body ground state $\ket{\Psi_\mathrm{GS}}$ and indices $i,j$ are restricted to be within the subsystem A. In Fig. \ref{fig:Fig2}, entanglement spectra for both lattices (a) before and (b) after phase transitions are shown. In both figures, the spectral flow can be seen, which is a characteristic feature of nontrivial topology of the band.

\section{Phase transitions in non-interacting Chern insulators}
We investigate topological phase transitions using the entanglement entropy. Entanglement entropy is calculated from the correlation matrix using a relation
		\begin{equation}
			S_{A} = -\sum_i \left(\zeta_i \ln{(\zeta_i)}+(1-\zeta_i)\ln{(1-\zeta_i)} \right),
			\label{eq:Entr}
		\end{equation}
where $\zeta_i$ are the eigenvalues of the correlation matrix $C_{ij}$. From Eq. (\ref{eq:Entr}), one can clearly see that eigenvalues $\zeta_i=0$ or $\zeta_i=1$ have no contribution to the single-particle entanglement entropy. A general feature of a single-particle entanglement entropy for a system in a topologically nontrivial phase is that it cannot be transformed to a system with zero entanglement without closing the energy gap. This situation holds if the entanglement spectrum is gapless, observed in Fig. \ref{fig:Fig2}(a) and (b). 

In order to identify topological phase transition points, the single-particle entanglement entropies as a function of $t$ are analyzed in Fig. \ref{fig:Fig2}(c). For the Lieb lattice, the entanglement entropy (marked with black squares) monotonically increases with an increase of $t$ from $t=0$ to some finite value. At $t\sim 1.4$, a drop of the entanglement entropy is seen, which coincides with a topological phase transition, indicated by the red vertical line in Fig. \ref{fig:Fig1}(c). Similarly, a drop of the entanglement entropy is observed at the phase transition on kagome lattice (denoted by red triangles) at $t\sim 0.55$. Thus, the entanglement entropy properly identified the point of a topological phase transition. At these points, the divergence of a standard deviation of Berry curvature of the lowest band is also seen in Fig. \ref{fig:Fig2}(d) for both lattices, but has a sharper character for the Lieb lattice, increasing up to $\sigma (F_{12})\sim 6$ in comparison to $\sigma (F_{12})\sim 1.5$ for kagome lattice. For both lattices, the Berry curvature is approximately uniform over the entire Brillouin zone for all values of $t$, except for the vicinity of the phase transitions. An increase of the Berry curvature dispersion around $t=0$ fo the Lieb lattice is related to a topological phase transition that occurs for infinitesimally small $t$, between the isolated sites forming the middle band with $C=0$, and the band with $C=-2$, when these sites are coupled to a checkerboard lattice. One can notice also that Berry curvature dispersion is smaller for kagome lattice in comparison to the Lieb lattice. Flatness of the Berry curvature was shown as one of main factors responsible for stabilization of FCI phases \cite{BerryParameswaran, Zoology, We}.

\section{Phase transition at $1/3$ filling}
Many-body effects are studied by adding density-density interaction term in the form
    \begin{equation}
    V=V_{\mathrm{nn}}\sum_{\braket{i,j}}n_in_j+V_{\mathrm{nnn}}\sum_{\braket{\braket{i,j}}}n_in_j,
\label{eq:Vint}
    \end{equation}
where $n_i$ is a density operator on the site $i$, $V_{\mathrm{nn}}$ denotes an interaction between the nearest neighbors and $V_{\mathrm{nnn}}$ between the next-nearest neighbors, see Fig. \ref{fig:Fig1}. Correlation effects are examined within the lowest band filled by spinless fermions for $\nu = 1/3$ filling factor, as it can lead to appearance of Laughlin-like FCI phase. Firstly, we perform calculations within the flat-band approximation, neglecting the kinetic energies and the presence of higher bands. A validity of this approximation will be discussed in details at the end of this Section. All calculations are performed for finite size $N_x\times N_y$ samples in a torus geometry, where $N_x$ ($N_y$) is a number of unit cells in $x$ ($y$) direction. Due to the translational symmetry and the momentum conservation of the two-particle Coulomb scattering term, many-body eigenstates can be labeled by a total momentum quantum numbers $K_x$ and $K_y$, which are the sum of the momentum quantum numbers of each of the $N$ particles modulo $N_x$ and $N_y$. As a representative sample, we take $N_x\times N_y=4\times 6$ system with $N=8$ particles. 

\subsection{The many-body energy spectrum}
\begin{figure}
	\centerline{\includegraphics[width=\columnwidth]{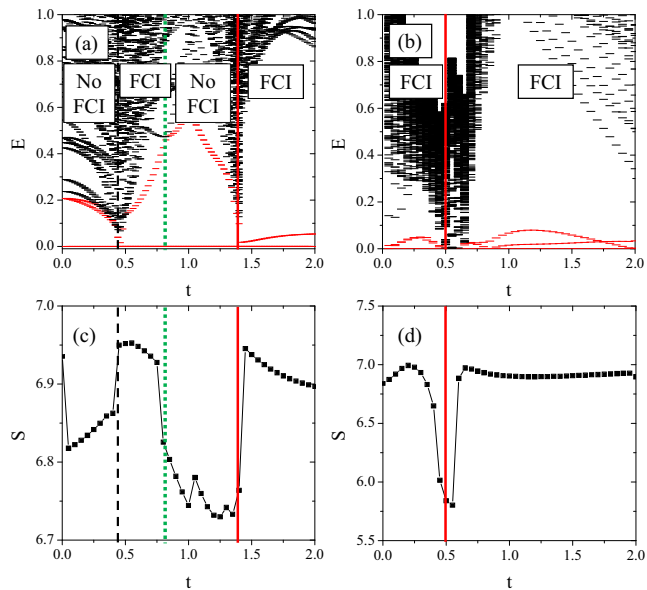}}
		\caption{Evolution of many-body energy spectrum of the lowest band for $\nu = 1/3$ filling factor for the nearest neighbors interaction term $V_{\mathrm{nn}} = 15$ and the next-nearest neighbors interaction $V_{\mathrm{nnn}} = 10$ as a function of parameter $t$ for (a) the Lieb lattice and (b) kagome lattice. Three lowest energy states are indicated by a red color. Parameters regions with and without FCI phases are labeled and separated by vertical lines. Many-body energy gap closing points are indicated by the black dash line and the red solid lines. (c) and (d) Entanglement entropies calculated from three lowest energy states for the Lieb and kagome lattices.} 
		\label{fig:Fig3}
\end{figure}
The many-body energy spectrum as a function of hopping integral $t$ for the nearest-neighbors interaction $V_{\mathrm{nn}}=15$ and next-nearest neighbor interaction $V_{\mathrm{nnn}} = 10$ is analyzed in Fig. \ref{fig:Fig3} (a) and (b) for the Lieb and kagome lattices, respectively. Interacting parameters exceed bands dispersions, hence the influence of other band has to be taken into account, which will be discussed later. However, as long as the calculations are performed within the flat band limit, only the relative magnitudes of parameters $V_{\mathrm{nn}}, \, V_{\mathrm{nnn}}$ are important. As we focus on the $1/3$ filling and possible existence of FCI phase, we indicate the three lowest energy states by a red color. They are separated by a well-defined many-body energy gap from the excited states for $t>1.4$ for the Lieb lattice, and in the range of $0<t<0.5$ and $t>0.7$ for kagome lattice. In these cases, we confirm that the system is in FCI Laughlin-like phase by looking at its characteristic features: appropriate ground state momenta $(N_x, \, N_y)=(0, \, 0),(0, \, 2),(0, \, 4)$, spectral flow, and entanglement spectrum (not shown here). We also notice FCI phase for the Lieb lattice in the range of $0.5<t<0.7$. At $t\sim 0.7$, one of the three lowest energy states intersects with an excited state, indicated by a green dotted line, and for larger $t$ FCI phase is not stable anymore. At $t\sim 1.4$ for the Lieb lattice (indicated by a red vertical line), the many-body energy gap closes, which coincides with the single-particle energy gap closure and the topological phase transition within a non-interacting model. At $t\sim 0.5$, also many-body energy gap closing is observed (marked by a black dash line), which suggests other phase transition. We look at results for different system sizes, $N_x\times N_y=3\times 5$, $N_x\times N_y=3\times 6$, $N_x\times N_y=5\times 6$ and also several interaction strengths. We observe a gap closing at $t\sim 0.5$ (for $V_{\mathrm{nn}}>6$) and always at $t\sim 1.4$. On the other hand, we notice sensitivity of the FCI phase to the plaquette sizes in the region $0.5<t<1.4$. Thus, observation of FCI phase is affected by finite size effects, but the phase transitions at $t\sim 0.5$ and $t\sim 1.4$ are always present. 

We perform similar analysis for the kagome lattice, with evolution of energy spectrum shown in Fig. \ref{fig:Fig3}(b). Here, the many-body energy gap closes around $t\sim 0.5$, indicated by a red vertical line, and reopens again around $t\sim 0.7$. This phase transition spreads over a region with several $t$ values and can be related to smoother behavior of the Berry curvature at bands intersection at $t=0.55$, in contrast to the sharp peak at $t=1.4$ for the Lieb lattice, see Fig. \ref{fig:Fig2}(d). 

We note that for both lattices we have also looked at different filling factor with $N=5,6,7,9,10$ particles on $N_x\times N_y=4\times 6$ system, however we have not observed closing of the energy gap at $t=1.4$ for the Lieb lattice and at $t=0.55$ for the kagome lattice. Thus, topological phase transitions observed for an interacting system correspond to the disappearance of FCI phases, which can be related to the divergence of the Berry curvature at these phase transition points. 

\subsection{The particle entanglement entropy}
FCI and other unidentified phases with phase transitions indicated by vertical lines in Fig. \ref{fig:Fig3}(a) and (b) are analyzed in Fig. \ref{fig:Fig3}(c) and (d) using the particle entanglement entropy $S$. This type of measure of entanglement is especially sensitive to the quantum statistics \cite{Zoology, Zozulya1, Zozulya2}. The system is divided into $n_A=3$ particles of the state with traced out the degrees of freedom carried by the remaining $n_B = 5$ particles. We form the density matrix from the three lowest states, as a sum with equal weights, $\rho=\sum_i 1/3|\Psi_i\rangle\langle\Psi_i|$, with $i=1,2,3$. In the case of FCI these are the three quasidegenerate states forming the ground state manifold. The diagonalized reduced density matrix of $n_A=3$ particles, $\rho_A=Tr_B \rho$ gives eigenvalues $\zeta=e^{-\xi}$ which defines the particle entanglement energies $\xi$. The entanglement entropy is obtained directly from the entanglement spectrum
\begin{equation}
S_A =-\sum_{i}\zeta_i \ln(\zeta_i),
\end{equation}
where the sum runs over all the entanglement energies.

The upper bound of the entanglement entropy is given by the logarithm of the maximum number of equal nonzero eigenvalues, $S_{\mathrm{max}}=\ln \binom{N_{orb}}{n_a}$, with $N_{orb}=N_x\cdot N_y$ being the number of single-particle orbitals within the band.  In our case $S_{\mathrm{max}}=\ln \binom{24}{3}\approx 7.61$, which should be further reduced due to generalized Pauli principle \cite{Zoology, Zozulya1, Zozulya2}. Thus, the correlations in the system reduce entanglement entropy from its maximal value. The lower bound is related to the antisymmetrization of many-body wave function. For a state described by a single Slater determinant (and also for equal weight superposition of three-fold degenerate ground state) is given by $S_{\mathrm{min}}=\ln \binom{N}{n_A}$, in our case $\ln \binom{8}{3}\approx 4.02$. The excess of particle entanglement over $S_{\mathrm{min}}$ can show the number of Slater determinant terms of similar amplitudes that need to be combined to produce the ground state.  

The entanglement entropy has the highest value within regions with FCI phases, a bit larger in the case of kagome lattice, around $S_A\sim 7.0$, compare Fig. \ref{fig:Fig3}(c) and (d). Obtained values are smaller than the values of the entanglement entropy expected for $1/3$ Laughlin states in the fractional quantum Hall effect, given by a relation $S_{1/3}=\ln \binom{N}{n_A} + \alpha(n_A)\approx7.31$, with  $\alpha(n_A)=n_A \ln 3$ \cite{Zozulya2}. Transitions to other unidentified phases, labeled as "No FCI'', coincide with a drop of the entanglement entropy. This occurs at $t\sim 0.5$ and $t\sim 1.4$ denoted by black and red solid lines, where the many-body energy gap closes, and at $t\sim 0.7$ for the Lieb lattice, indicated by a green dotted line, where one of the ground states intersect with an excited state. Similarly, at topological phase transition point around $t\sim 0.55$ for kagome lattice a rapid drop of the entanglement entropy is observed. 

\begin{figure}
	\centerline{\includegraphics[width=\columnwidth]{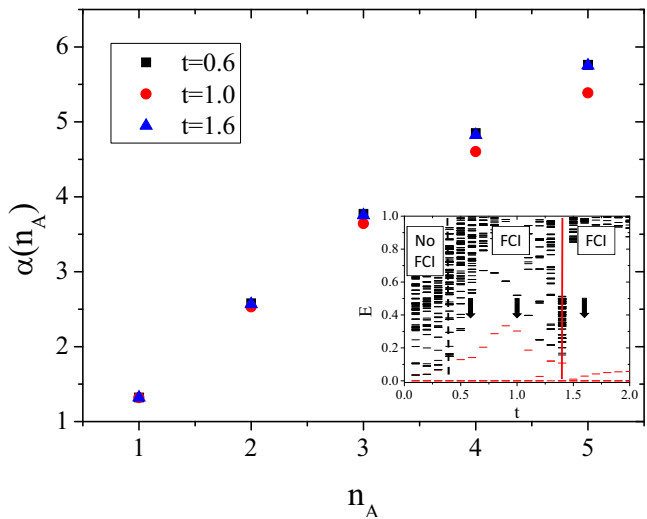}}
	\caption{Scaling of the particle entanglement entropy $\alpha (n_A)$ term with a number of particles in a subsystem A at $t=0.6$, $t=1.0$, and $t=1.6$ for $N_x\times N_y=5\times 6$ system. Energy spectrum as a function of $t$ is shown in the inset, with three values of the hopping integral $t$ indicated by black arrows, taken for results from the main panel.}
	 \label{fig:Fig4}
\end{figure}
In Fig. \ref{fig:Fig4} we plot $\alpha (n_A)$ term as a function of a number of particles in a subsystem A for  $N_x\times N_y=5\times 6$. A linear scaling is expected for Laughlin-like phases in a limit of the large number of particles \cite{Zozulya1, Zozulya2}. For FCI phase, a quasilinear dependence is observed at $t=0.6$ and $t=1.6$, and slight sublinear dependence at $t=1.0$. In the last case, spread of the three-fold degenerate ground states in the energy spectrum is the largest (see the inset in Fig. \ref{fig:Fig4}), which suggests a less stable phase. Linear regression for $t=0.6$ and $t=1.6$ yields slope 1.02, which is close to $\ln 3 \approx 1.1$. We note that for smaller system $N_x\times N_y=4\times 6$ a sublinear dependence is observed for FCI phase (not shown here), which may be a result of a too small number of particles, as the linear scaling is expected in a limit of large number of particles. 

\subsection{The higher band effect}
\begin{figure}
	\centerline{\includegraphics[width=\columnwidth]{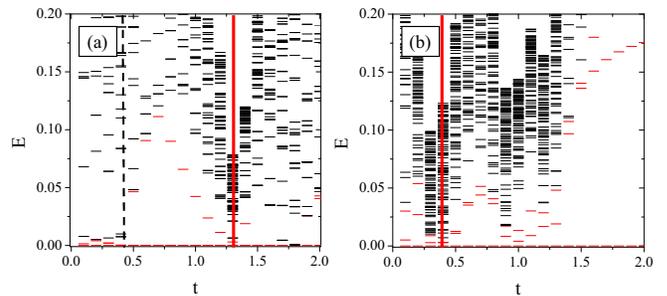}}
	\caption{Many-body energy spectra for calculations including two lowest bands (48 states in total) with $N=8$ particles (filling factor of the lowest band $\nu = 1/3$) for the next-nearest neighbors interaction term $V_{\mathrm{nn}} = 15$ and the next-nearest neighbors interaction term $V_{\mathrm{nnn}} = 10$ in the case of (a) the Lieb lattice and (b) kagome lattice. The system size is $N_x\times N_y=4\times 6$. In (a), closing of the energy gap at $t\sim 0.4$ and at $t\sim 1.3$ are observed, indicated by black dash and red solid lines, respectively. In (b), closing of the energy gap at $t\sim 0.4$ is observed and marked by a red solid line.}
	 \label{fig:Fig5}
\end{figure}
In general, the flat band limit approximation we have used is justified when the energy gap $E_{\mathrm{gap}}$,  the interaction energy scale $V$, and the band dispersion $E_d$ fulfill the condition $E_g \gg V \gg E_d$. In considered models, the energy bands are dispersive and the energy gap between the two lowest bands is not well-defined in a case of the Lieb lattice, Fig. \ref{fig:Fig1}(c) and (d). The flat band limit condition is not satisfied, however, as was shown before by Neupert et al. in Ref. \cite{FarExceed, SingleParticleGrushin}, FCI phases can be stable for an interaction far exceeding the band gap. We analyze this problem for the models used in this work.

In Fig. \ref{fig:Fig5} the many-body energy spectra as a function of parameter $t$ calculated with two lowest bands included are presented. There are 48 states in total with $N=8$ particles, which corresponds to filling factor of the lowest band $\nu = 1/3$. These results can be compared with Fig. \ref{fig:Fig3}(a) and (b), where only the lowest band is taken into account. One can notice that for both lattices the energy scales are five times smaller. The three-fold ground state manifolds are depicted by red bars, but for most of $t$ values they are not clearly separated from the excited states. 
Energy gap opening is seen after $t\sim 0.4$ indicated by a black dashed line and the gap closings at $t\sim 1.3$ for the Lieb lattice, and after $t\sim 0.4$ for kagome lattice (red solid vertical lines). Thus, phase transitions observed for results within a flat band limit are still present, but slightly shifted energetically to smaller values. We have verified that for $0.7<t<1.2$ and $t=1.4$ for Lieb lattice, and in the range  $0.9<t<1.3$ for kagome lattice, the ground state momenta agree with momenta of FCI phase. Thus, FCI phase is stable for some values of parameters but outside this region its stability is sensitive to interaction with the higher band and also to a specific choice of model parameters (different interaction parameters may restore FCI phase). However, the phase transitions can be expected to have an universal character.

\section{Conclusions}
We have studied topological phase transitions on the Lieb and kagome lattices driven by a single-particle hopping integral by analyzing the low energy spectra, the entanglement spectra and the entanglement entropy. We considered a non-interacting case of the empty band and with the many-body effects of partially filled lowest energy bands. Nontrivial topology of bands was determined by directly calculating Chern numbers, and also by observation of the spectral flow in the entanglement spectra. At the single-particle level, we have found the topological phase transition between the two lowest bands for hopping integral $t\sim 1.4$ with Chern number change from $C=1$ to $C=-1$ for the Lieb lattice and at $t\sim 0.55$ with Chern number change from $C=-1$ to $C=1$ for kagome lattice. During topological phase transition, a rapid decrease of the single-particle entanglement entropy and divergence of the Berry curvature are seen. At filling factor $1/3$ for the lowest band, FCI phase was recognized and distinguished from other unidentified phases. The many-body energy gap closing points were observed to coincide with topological phase transition points associated with a change in the Chern number. We have related it to the disappearance of FCI phases due to divergence of the Berry curvature, as for different filling factors no gap closure was noticed. We have also found that the particle entanglement entropy of FCI phase have always the largest value in comparison to other unidentified phases. A quasilinear behavior of the entanglement entropy $\alpha(n_A)$ term is observed for FCI phase, similarly to linear dependence expected for Laughlin-like phase in FQHE. The presence of FCI phases in a certain parameter range is sensitive to the finite size effects and interaction with higher bands, however topological phase transitions are shown to occur for arbitrary system sizes, thus are expected to be observed in the thermodynamic limit. 
\begin{acknowledgments}
The authors acknowledge partial financial support from National Science Center (NCN), Poland, grant Maestro No. 2014/14/A/ST3/00654. Our calculations were performed in the Wroc\l{}aw Center for Networking and Supercomputing. 
\end{acknowledgments}
\bibliography{lieb_tpt}

\begin{thebibliography}{56}%
\makeatletter
\providecommand \@ifxundefined [1]{%
 \@ifx{#1\undefined}
}%
\providecommand \@ifnum [1]{%
 \ifnum #1\expandafter \@firstoftwo
 \else \expandafter \@secondoftwo
 \fi
}%
\providecommand \@ifx [1]{%
 \ifx #1\expandafter \@firstoftwo
 \else \expandafter \@secondoftwo
 \fi
}%
\providecommand \natexlab [1]{#1}%
\providecommand \enquote  [1]{``#1''}%
\providecommand \bibnamefont  [1]{#1}%
\providecommand \bibfnamefont [1]{#1}%
\providecommand \citenamefont [1]{#1}%
\providecommand \href@noop [0]{\@secondoftwo}%
\providecommand \href [0]{\begingroup \@sanitize@url \@href}%
\providecommand \@href[1]{\@@startlink{#1}\@@href}%
\providecommand \@@href[1]{\endgroup#1\@@endlink}%
\providecommand \@sanitize@url [0]{\catcode `\\12\catcode `\$12\catcode
  `\&12\catcode `\#12\catcode `\^12\catcode `\_12\catcode `\%12\relax}%
\providecommand \@@startlink[1]{}%
\providecommand \@@endlink[0]{}%
\providecommand \url  [0]{\begingroup\@sanitize@url \@url }%
\providecommand \@url [1]{\endgroup\@href {#1}{\urlprefix }}%
\providecommand \urlprefix  [0]{URL }%
\providecommand \Eprint [0]{\href }%
\providecommand \doibase [0]{http://dx.doi.org/}%
\providecommand \selectlanguage [0]{\@gobble}%
\providecommand \bibinfo  [0]{\@secondoftwo}%
\providecommand \bibfield  [0]{\@secondoftwo}%
\providecommand \translation [1]{[#1]}%
\providecommand \BibitemOpen [0]{}%
\providecommand \bibitemStop [0]{}%
\providecommand \bibitemNoStop [0]{.\EOS\space}%
\providecommand \EOS [0]{\spacefactor3000\relax}%
\providecommand \BibitemShut  [1]{\csname bibitem#1\endcsname}%
\let\auto@bib@innerbib\@empty
\bibitem [{\citenamefont {Sun}\ \emph {et~al.}(2011)\citenamefont {Sun},
  \citenamefont {Gu}, \citenamefont {Katsura},\ and\ \citenamefont
  {Das~Sarma}}]{Sun}%
  \BibitemOpen
  \bibfield  {author} {\bibinfo {author} {\bibfnamefont {K.}~\bibnamefont
  {Sun}}, \bibinfo {author} {\bibfnamefont {Z.}~\bibnamefont {Gu}}, \bibinfo
  {author} {\bibfnamefont {H.}~\bibnamefont {Katsura}}, \ and\ \bibinfo
  {author} {\bibfnamefont {S.}~\bibnamefont {Das~Sarma}},\ }\href {\doibase
  10.1103/PhysRevLett.106.236803} {\bibfield  {journal} {\bibinfo  {journal}
  {Phys. Rev. Lett.}\ }\textbf {\bibinfo {volume} {106}},\ \bibinfo {pages}
  {236803} (\bibinfo {year} {2011})}\BibitemShut {NoStop}%
\bibitem [{\citenamefont {Neupert}\ \emph {et~al.}(2011)\citenamefont
  {Neupert}, \citenamefont {Santos}, \citenamefont {Chamon},\ and\
  \citenamefont {Mudry}}]{Neupert}%
  \BibitemOpen
  \bibfield  {author} {\bibinfo {author} {\bibfnamefont {T.}~\bibnamefont
  {Neupert}}, \bibinfo {author} {\bibfnamefont {L.}~\bibnamefont {Santos}},
  \bibinfo {author} {\bibfnamefont {C.}~\bibnamefont {Chamon}}, \ and\ \bibinfo
  {author} {\bibfnamefont {C.}~\bibnamefont {Mudry}},\ }\href {\doibase
  10.1103/PhysRevLett.106.236804} {\bibfield  {journal} {\bibinfo  {journal}
  {Phys. Rev. Lett.}\ }\textbf {\bibinfo {volume} {106}},\ \bibinfo {pages}
  {236804} (\bibinfo {year} {2011})}\BibitemShut {NoStop}%
\bibitem [{\citenamefont {Tang}\ \emph {et~al.}(2011)\citenamefont {Tang},
  \citenamefont {Mei},\ and\ \citenamefont {Wen}}]{Tang}%
  \BibitemOpen
  \bibfield  {author} {\bibinfo {author} {\bibfnamefont {E.}~\bibnamefont
  {Tang}}, \bibinfo {author} {\bibfnamefont {J.-W.}\ \bibnamefont {Mei}}, \
  and\ \bibinfo {author} {\bibfnamefont {X.-G.}\ \bibnamefont {Wen}},\ }\href
  {\doibase 10.1103/PhysRevLett.106.236802} {\bibfield  {journal} {\bibinfo
  {journal} {Phys. Rev. Lett.}\ }\textbf {\bibinfo {volume} {106}},\ \bibinfo
  {pages} {236802} (\bibinfo {year} {2011})}\BibitemShut {NoStop}%
\bibitem [{\citenamefont {Chang}\ \emph {et~al.}(2013)\citenamefont {Chang},
  \citenamefont {Zhang}, \citenamefont {Feng}, \citenamefont {Shen},
  \citenamefont {Zhang}, \citenamefont {Guo}, \citenamefont {Li}, \citenamefont
  {Ou}, \citenamefont {Wei}, \citenamefont {Wang}, \citenamefont {Ji},
  \citenamefont {Feng}, \citenamefont {Ji}, \citenamefont {Chen}, \citenamefont
  {Jia}, \citenamefont {Dai}, \citenamefont {Fang}, \citenamefont {Zhang},
  \citenamefont {He}, \citenamefont {Wang}, \citenamefont {Lu}, \citenamefont
  {Ma},\ and\ \citenamefont {Xue}}]{ChernExperiment}%
  \BibitemOpen
  \bibfield  {author} {\bibinfo {author} {\bibfnamefont {C.-Z.}\ \bibnamefont
  {Chang}}, \bibinfo {author} {\bibfnamefont {J.}~\bibnamefont {Zhang}},
  \bibinfo {author} {\bibfnamefont {X.}~\bibnamefont {Feng}}, \bibinfo {author}
  {\bibfnamefont {J.}~\bibnamefont {Shen}}, \bibinfo {author} {\bibfnamefont
  {Z.}~\bibnamefont {Zhang}}, \bibinfo {author} {\bibfnamefont
  {M.}~\bibnamefont {Guo}}, \bibinfo {author} {\bibfnamefont {K.}~\bibnamefont
  {Li}}, \bibinfo {author} {\bibfnamefont {Y.}~\bibnamefont {Ou}}, \bibinfo
  {author} {\bibfnamefont {P.}~\bibnamefont {Wei}}, \bibinfo {author}
  {\bibfnamefont {L.-L.}\ \bibnamefont {Wang}}, \bibinfo {author}
  {\bibfnamefont {Z.-Q.}\ \bibnamefont {Ji}}, \bibinfo {author} {\bibfnamefont
  {Y.}~\bibnamefont {Feng}}, \bibinfo {author} {\bibfnamefont {S.}~\bibnamefont
  {Ji}}, \bibinfo {author} {\bibfnamefont {X.}~\bibnamefont {Chen}}, \bibinfo
  {author} {\bibfnamefont {J.}~\bibnamefont {Jia}}, \bibinfo {author}
  {\bibfnamefont {X.}~\bibnamefont {Dai}}, \bibinfo {author} {\bibfnamefont
  {Z.}~\bibnamefont {Fang}}, \bibinfo {author} {\bibfnamefont {S.-C.}\
  \bibnamefont {Zhang}}, \bibinfo {author} {\bibfnamefont {K.}~\bibnamefont
  {He}}, \bibinfo {author} {\bibfnamefont {Y.}~\bibnamefont {Wang}}, \bibinfo
  {author} {\bibfnamefont {L.}~\bibnamefont {Lu}}, \bibinfo {author}
  {\bibfnamefont {X.-C.}\ \bibnamefont {Ma}}, \ and\ \bibinfo {author}
  {\bibfnamefont {Q.-K.}\ \bibnamefont {Xue}},\ }\href {\doibase
  10.1126/science.1234414} {\bibfield  {journal} {\bibinfo  {journal}
  {Science}\ }\textbf {\bibinfo {volume} {340}},\ \bibinfo {pages} {167}
  (\bibinfo {year} {2013})}\BibitemShut {NoStop}%
\bibitem [{\citenamefont {Jotzu}\ \emph {et~al.}(2014)\citenamefont {Jotzu},
  \citenamefont {Messer}, \citenamefont {Desbuquois}, \citenamefont {Lebrat},
  \citenamefont {Uehlinger}, \citenamefont {Greif},\ and\ \citenamefont
  {Esslinger}}]{ChernExperiment2}%
  \BibitemOpen
  \bibfield  {author} {\bibinfo {author} {\bibfnamefont {G.}~\bibnamefont
  {Jotzu}}, \bibinfo {author} {\bibfnamefont {M.}~\bibnamefont {Messer}},
  \bibinfo {author} {\bibfnamefont {R.}~\bibnamefont {Desbuquois}}, \bibinfo
  {author} {\bibfnamefont {M.}~\bibnamefont {Lebrat}}, \bibinfo {author}
  {\bibfnamefont {T.}~\bibnamefont {Uehlinger}}, \bibinfo {author}
  {\bibfnamefont {D.}~\bibnamefont {Greif}}, \ and\ \bibinfo {author}
  {\bibfnamefont {T.}~\bibnamefont {Esslinger}},\ }\href
  {http://dx.doi.org/10.1038/nature13915} {\bibfield  {journal} {\bibinfo
  {journal} {Nature}\ }\textbf {\bibinfo {volume} {515}},\ \bibinfo {pages}
  {237} (\bibinfo {year} {2014})},\ \bibinfo {note} {letter}\BibitemShut
  {NoStop}%
\bibitem [{\citenamefont {Haldane}(1988)}]{Haldane}%
  \BibitemOpen
  \bibfield  {author} {\bibinfo {author} {\bibfnamefont {F.~D.~M.}\
  \bibnamefont {Haldane}},\ }\href {\doibase 10.1103/PhysRevLett.61.2015}
  {\bibfield  {journal} {\bibinfo  {journal} {Phys. Rev. Lett.}\ }\textbf
  {\bibinfo {volume} {61}},\ \bibinfo {pages} {2015} (\bibinfo {year}
  {1988})}\BibitemShut {NoStop}%
\bibitem [{\citenamefont {Liu}\ \emph {et~al.}(2012{\natexlab{a}})\citenamefont
  {Liu}, \citenamefont {Chen}, \citenamefont {Wang},\ and\ \citenamefont
  {Gong}}]{flatkagome}%
  \BibitemOpen
  \bibfield  {author} {\bibinfo {author} {\bibfnamefont {R.}~\bibnamefont
  {Liu}}, \bibinfo {author} {\bibfnamefont {W.-C.}\ \bibnamefont {Chen}},
  \bibinfo {author} {\bibfnamefont {Y.-F.}\ \bibnamefont {Wang}}, \ and\
  \bibinfo {author} {\bibfnamefont {C.-D.}\ \bibnamefont {Gong}},\ }\href
  {http://stacks.iop.org/0953-8984/24/i=30/a=305602} {\bibfield  {journal}
  {\bibinfo  {journal} {Journal of Physics: Condensed Matter}\ }\textbf
  {\bibinfo {volume} {24}},\ \bibinfo {pages} {305602} (\bibinfo {year}
  {2012}{\natexlab{a}})}\BibitemShut {NoStop}%
\bibitem [{\citenamefont {Wu}\ \emph {et~al.}(2012)\citenamefont {Wu},
  \citenamefont {Bernevig},\ and\ \citenamefont {Regnault}}]{Zoology}%
  \BibitemOpen
  \bibfield  {author} {\bibinfo {author} {\bibfnamefont {Y.-L.}\ \bibnamefont
  {Wu}}, \bibinfo {author} {\bibfnamefont {B.~A.}\ \bibnamefont {Bernevig}}, \
  and\ \bibinfo {author} {\bibfnamefont {N.}~\bibnamefont {Regnault}},\ }\href
  {\doibase 10.1103/PhysRevB.85.075116} {\bibfield  {journal} {\bibinfo
  {journal} {Phys. Rev. B}\ }\textbf {\bibinfo {volume} {85}},\ \bibinfo
  {pages} {075116} (\bibinfo {year} {2012})}\BibitemShut {NoStop}%
\bibitem [{\citenamefont {Jaworowski}\ \emph {et~al.}(2015)\citenamefont
  {Jaworowski}, \citenamefont {Manolescu},\ and\ \citenamefont {Potasz}}]{We}%
  \BibitemOpen
  \bibfield  {author} {\bibinfo {author} {\bibfnamefont {B.}~\bibnamefont
  {Jaworowski}}, \bibinfo {author} {\bibfnamefont {A.}~\bibnamefont
  {Manolescu}}, \ and\ \bibinfo {author} {\bibfnamefont {P.}~\bibnamefont
  {Potasz}},\ }\href@noop {} {\bibfield  {journal} {\bibinfo  {journal}
  {Physical Review B}\ }\textbf {\bibinfo {volume} {92}},\ \bibinfo {pages}
  {245119} (\bibinfo {year} {2015})}\BibitemShut {NoStop}%
\bibitem [{\citenamefont {Trescher}\ and\ \citenamefont
  {Bergholtz}(2012)}]{trescher2012flat}%
  \BibitemOpen
  \bibfield  {author} {\bibinfo {author} {\bibfnamefont {M.}~\bibnamefont
  {Trescher}}\ and\ \bibinfo {author} {\bibfnamefont {E.~J.}\ \bibnamefont
  {Bergholtz}},\ }\href@noop {} {\bibfield  {journal} {\bibinfo  {journal}
  {Physical Review B}\ }\textbf {\bibinfo {volume} {86}},\ \bibinfo {pages}
  {241111} (\bibinfo {year} {2012})}\BibitemShut {NoStop}%
\bibitem [{\citenamefont {Wang}\ and\ \citenamefont
  {Ran}(2011)}]{wang2011dice}%
  \BibitemOpen
  \bibfield  {author} {\bibinfo {author} {\bibfnamefont {F.}~\bibnamefont
  {Wang}}\ and\ \bibinfo {author} {\bibfnamefont {Y.}~\bibnamefont {Ran}},\
  }\href@noop {} {\bibfield  {journal} {\bibinfo  {journal} {Physical Review
  B}\ }\textbf {\bibinfo {volume} {84}},\ \bibinfo {pages} {241103} (\bibinfo
  {year} {2011})}\BibitemShut {NoStop}%
\bibitem [{\citenamefont {Chen}\ \emph {et~al.}(2012)\citenamefont {Chen},
  \citenamefont {Liu}, \citenamefont {Wang},\ and\ \citenamefont
  {Gong}}]{flatstar}%
  \BibitemOpen
  \bibfield  {author} {\bibinfo {author} {\bibfnamefont {W.-C.}\ \bibnamefont
  {Chen}}, \bibinfo {author} {\bibfnamefont {R.}~\bibnamefont {Liu}}, \bibinfo
  {author} {\bibfnamefont {Y.-F.}\ \bibnamefont {Wang}}, \ and\ \bibinfo
  {author} {\bibfnamefont {C.-D.}\ \bibnamefont {Gong}},\ }\href {\doibase
  10.1103/PhysRevB.86.085311} {\bibfield  {journal} {\bibinfo  {journal} {Phys.
  Rev. B}\ }\textbf {\bibinfo {volume} {86}},\ \bibinfo {pages} {085311}
  (\bibinfo {year} {2012})}\BibitemShut {NoStop}%
\bibitem [{\citenamefont {Liu}\ \emph {et~al.}(2013{\natexlab{a}})\citenamefont
  {Liu}, \citenamefont {Chen}, \citenamefont {Wang},\ and\ \citenamefont
  {Gong}}]{flatsqoct}%
  \BibitemOpen
  \bibfield  {author} {\bibinfo {author} {\bibfnamefont {X.-P.}\ \bibnamefont
  {Liu}}, \bibinfo {author} {\bibfnamefont {W.-C.}\ \bibnamefont {Chen}},
  \bibinfo {author} {\bibfnamefont {Y.-F.}\ \bibnamefont {Wang}}, \ and\
  \bibinfo {author} {\bibfnamefont {C.-D.}\ \bibnamefont {Gong}},\ }\href
  {http://stacks.iop.org/0953-8984/25/i=30/a=305602} {\bibfield  {journal}
  {\bibinfo  {journal} {Journal of Physics: Condensed Matter}\ }\textbf
  {\bibinfo {volume} {25}},\ \bibinfo {pages} {305602} (\bibinfo {year}
  {2013}{\natexlab{a}})}\BibitemShut {NoStop}%
\bibitem [{\citenamefont {Tsui}\ \emph {et~al.}(1982)\citenamefont {Tsui},
  \citenamefont {Stormer},\ and\ \citenamefont {Gossard}}]{tsui}%
  \BibitemOpen
  \bibfield  {author} {\bibinfo {author} {\bibfnamefont {D.~C.}\ \bibnamefont
  {Tsui}}, \bibinfo {author} {\bibfnamefont {H.~L.}\ \bibnamefont {Stormer}}, \
  and\ \bibinfo {author} {\bibfnamefont {A.~C.}\ \bibnamefont {Gossard}},\
  }\href {\doibase 10.1103/PhysRevLett.48.1559} {\bibfield  {journal} {\bibinfo
   {journal} {Phys. Rev. Lett.}\ }\textbf {\bibinfo {volume} {48}},\ \bibinfo
  {pages} {1559} (\bibinfo {year} {1982})}\BibitemShut {NoStop}%
\bibitem [{\citenamefont {Laughlin}(1983)}]{laughlin}%
  \BibitemOpen
  \bibfield  {author} {\bibinfo {author} {\bibfnamefont {R.~B.}\ \bibnamefont
  {Laughlin}},\ }\href {\doibase 10.1103/PhysRevLett.50.1395} {\bibfield
  {journal} {\bibinfo  {journal} {Phys. Rev. Lett.}\ }\textbf {\bibinfo
  {volume} {50}},\ \bibinfo {pages} {1395} (\bibinfo {year}
  {1983})}\BibitemShut {NoStop}%
\bibitem [{\citenamefont {Jain}(1989)}]{Jain}%
  \BibitemOpen
  \bibfield  {author} {\bibinfo {author} {\bibfnamefont {J.~K.}\ \bibnamefont
  {Jain}},\ }\href {\doibase 10.1103/PhysRevLett.63.199} {\bibfield  {journal}
  {\bibinfo  {journal} {Phys. Rev. Lett.}\ }\textbf {\bibinfo {volume} {63}},\
  \bibinfo {pages} {199} (\bibinfo {year} {1989})}\BibitemShut {NoStop}%
\bibitem [{\citenamefont {Sheng}\ \emph {et~al.}(2011)\citenamefont {Sheng},
  \citenamefont {Gu}, \citenamefont {Sun},\ and\ \citenamefont
  {Sheng}}]{SunNature}%
  \BibitemOpen
  \bibfield  {author} {\bibinfo {author} {\bibfnamefont {D.}~\bibnamefont
  {Sheng}}, \bibinfo {author} {\bibfnamefont {Z.-C.}\ \bibnamefont {Gu},
  \bibfnamefont {Gu}}, \bibinfo {author} {\bibfnamefont {K.}~\bibnamefont
  {Sun}}, \ and\ \bibinfo {author} {\bibfnamefont {L.}~\bibnamefont {Sheng}},\
  }\href {\doibase 10.1038/ncomms1380} {\bibfield  {journal} {\bibinfo
  {journal} {Nat Commun}\ }\textbf {\bibinfo {volume} {2}},\ \bibinfo {pages}
  {389} (\bibinfo {year} {2011})}\BibitemShut {NoStop}%
\bibitem [{\citenamefont {Regnault}\ and\ \citenamefont
  {Bernevig}(2011)}]{PRX}%
  \BibitemOpen
  \bibfield  {author} {\bibinfo {author} {\bibfnamefont {N.}~\bibnamefont
  {Regnault}}\ and\ \bibinfo {author} {\bibfnamefont {B.~A.}\ \bibnamefont
  {Bernevig}},\ }\href {\doibase 10.1103/PhysRevX.1.021014} {\bibfield
  {journal} {\bibinfo  {journal} {Phys. Rev. X}\ }\textbf {\bibinfo {volume}
  {1}},\ \bibinfo {pages} {021014} (\bibinfo {year} {2011})}\BibitemShut
  {NoStop}%
\bibitem [{\citenamefont {Kourtis}\ \emph {et~al.}(2012)\citenamefont
  {Kourtis}, \citenamefont {Venderbos},\ and\ \citenamefont
  {Daghofer}}]{KourtisTriangular}%
  \BibitemOpen
  \bibfield  {author} {\bibinfo {author} {\bibfnamefont {S.}~\bibnamefont
  {Kourtis}}, \bibinfo {author} {\bibfnamefont {J.~W.~F.}\ \bibnamefont
  {Venderbos}}, \ and\ \bibinfo {author} {\bibfnamefont {M.}~\bibnamefont
  {Daghofer}},\ }\href {\doibase 10.1103/PhysRevB.86.235118} {\bibfield
  {journal} {\bibinfo  {journal} {Phys. Rev. B}\ }\textbf {\bibinfo {volume}
  {86}},\ \bibinfo {pages} {235118} (\bibinfo {year} {2012})}\BibitemShut
  {NoStop}%
\bibitem [{\citenamefont {Yang}\ \emph
  {et~al.}(2012{\natexlab{a}})\citenamefont {Yang}, \citenamefont {Sun},\ and\
  \citenamefont {Das~Sarma}}]{YangDisorder}%
  \BibitemOpen
  \bibfield  {author} {\bibinfo {author} {\bibfnamefont {S.}~\bibnamefont
  {Yang}}, \bibinfo {author} {\bibfnamefont {K.}~\bibnamefont {Sun}}, \ and\
  \bibinfo {author} {\bibfnamefont {S.}~\bibnamefont {Das~Sarma}},\ }\href
  {\doibase 10.1103/PhysRevB.85.205124} {\bibfield  {journal} {\bibinfo
  {journal} {Phys. Rev. B}\ }\textbf {\bibinfo {volume} {85}},\ \bibinfo
  {pages} {205124} (\bibinfo {year} {2012}{\natexlab{a}})}\BibitemShut
  {NoStop}%
\bibitem [{\citenamefont {Liu}\ \emph {et~al.}(2012{\natexlab{b}})\citenamefont
  {Liu}, \citenamefont {Bergholtz}, \citenamefont {Fan},\ and\ \citenamefont
  {L\"auchli}}]{highchern1}%
  \BibitemOpen
  \bibfield  {author} {\bibinfo {author} {\bibfnamefont {Z.}~\bibnamefont
  {Liu}}, \bibinfo {author} {\bibfnamefont {E.~J.}\ \bibnamefont {Bergholtz}},
  \bibinfo {author} {\bibfnamefont {H.}~\bibnamefont {Fan}}, \ and\ \bibinfo
  {author} {\bibfnamefont {A.~M.}\ \bibnamefont {L\"auchli}},\ }\href {\doibase
  10.1103/PhysRevLett.109.186805} {\bibfield  {journal} {\bibinfo  {journal}
  {Phys. Rev. Lett.}\ }\textbf {\bibinfo {volume} {109}},\ \bibinfo {pages}
  {186805} (\bibinfo {year} {2012}{\natexlab{b}})}\BibitemShut {NoStop}%
\bibitem [{\citenamefont {Wang}\ \emph {et~al.}(2012)\citenamefont {Wang},
  \citenamefont {Yao}, \citenamefont {Gong},\ and\ \citenamefont
  {Sheng}}]{highchern2}%
  \BibitemOpen
  \bibfield  {author} {\bibinfo {author} {\bibfnamefont {Y.-F.}\ \bibnamefont
  {Wang}}, \bibinfo {author} {\bibfnamefont {H.}~\bibnamefont {Yao}}, \bibinfo
  {author} {\bibfnamefont {C.-D.}\ \bibnamefont {Gong}}, \ and\ \bibinfo
  {author} {\bibfnamefont {D.~N.}\ \bibnamefont {Sheng}},\ }\href {\doibase
  10.1103/PhysRevB.86.201101} {\bibfield  {journal} {\bibinfo  {journal} {Phys.
  Rev. B}\ }\textbf {\bibinfo {volume} {86}},\ \bibinfo {pages} {201101}
  (\bibinfo {year} {2012})}\BibitemShut {NoStop}%
\bibitem [{\citenamefont {Yang}\ \emph
  {et~al.}(2012{\natexlab{b}})\citenamefont {Yang}, \citenamefont {Gu},
  \citenamefont {Sun},\ and\ \citenamefont {Das~Sarma}}]{highchern3}%
  \BibitemOpen
  \bibfield  {author} {\bibinfo {author} {\bibfnamefont {S.}~\bibnamefont
  {Yang}}, \bibinfo {author} {\bibfnamefont {Z.-C.}\ \bibnamefont {Gu}},
  \bibinfo {author} {\bibfnamefont {K.}~\bibnamefont {Sun}}, \ and\ \bibinfo
  {author} {\bibfnamefont {S.}~\bibnamefont {Das~Sarma}},\ }\href {\doibase
  10.1103/PhysRevB.86.241112} {\bibfield  {journal} {\bibinfo  {journal} {Phys.
  Rev. B}\ }\textbf {\bibinfo {volume} {86}},\ \bibinfo {pages} {241112}
  (\bibinfo {year} {2012}{\natexlab{b}})}\BibitemShut {NoStop}%
\bibitem [{\citenamefont {Sterdyniak}\ \emph {et~al.}(2013)\citenamefont
  {Sterdyniak}, \citenamefont {Repellin}, \citenamefont {Bernevig},\ and\
  \citenamefont {Regnault}}]{highchern4}%
  \BibitemOpen
  \bibfield  {author} {\bibinfo {author} {\bibfnamefont {A.}~\bibnamefont
  {Sterdyniak}}, \bibinfo {author} {\bibfnamefont {C.}~\bibnamefont
  {Repellin}}, \bibinfo {author} {\bibfnamefont {B.~A.}\ \bibnamefont
  {Bernevig}}, \ and\ \bibinfo {author} {\bibfnamefont {N.}~\bibnamefont
  {Regnault}},\ }\href {\doibase 10.1103/PhysRevB.87.205137} {\bibfield
  {journal} {\bibinfo  {journal} {Phys. Rev. B}\ }\textbf {\bibinfo {volume}
  {87}},\ \bibinfo {pages} {205137} (\bibinfo {year} {2013})}\BibitemShut
  {NoStop}%
\bibitem [{\citenamefont {Cincio}\ and\ \citenamefont
  {Vidal}(2013)}]{cincio2013characterizing}%
  \BibitemOpen
  \bibfield  {author} {\bibinfo {author} {\bibfnamefont {L.}~\bibnamefont
  {Cincio}}\ and\ \bibinfo {author} {\bibfnamefont {G.}~\bibnamefont {Vidal}},\
  }\href@noop {} {\bibfield  {journal} {\bibinfo  {journal} {Physical review
  letters}\ }\textbf {\bibinfo {volume} {110}},\ \bibinfo {pages} {067208}
  (\bibinfo {year} {2013})}\BibitemShut {NoStop}%
\bibitem [{\citenamefont {Liu}\ \emph {et~al.}(2013{\natexlab{b}})\citenamefont
  {Liu}, \citenamefont {Kovrizhin},\ and\ \citenamefont
  {Bergholtz}}]{liu2013bulk}%
  \BibitemOpen
  \bibfield  {author} {\bibinfo {author} {\bibfnamefont {Z.}~\bibnamefont
  {Liu}}, \bibinfo {author} {\bibfnamefont {D.}~\bibnamefont {Kovrizhin}}, \
  and\ \bibinfo {author} {\bibfnamefont {E.~J.}\ \bibnamefont {Bergholtz}},\
  }\href@noop {} {\bibfield  {journal} {\bibinfo  {journal} {Physical Review
  B}\ }\textbf {\bibinfo {volume} {88}},\ \bibinfo {pages} {081106} (\bibinfo
  {year} {2013}{\natexlab{b}})}\BibitemShut {NoStop}%
\bibitem [{\citenamefont {Grushin}\ \emph {et~al.}(2015)\citenamefont
  {Grushin}, \citenamefont {Motruk}, \citenamefont {Zaletel},\ and\
  \citenamefont {Pollmann}}]{Johannes}%
  \BibitemOpen
  \bibfield  {author} {\bibinfo {author} {\bibfnamefont {A.~G.}\ \bibnamefont
  {Grushin}}, \bibinfo {author} {\bibfnamefont {J.}~\bibnamefont {Motruk}},
  \bibinfo {author} {\bibfnamefont {M.~P.}\ \bibnamefont {Zaletel}}, \ and\
  \bibinfo {author} {\bibfnamefont {F.}~\bibnamefont {Pollmann}},\ }\href@noop
  {} {\bibfield  {journal} {\bibinfo  {journal} {Physical Review B}\ }\textbf
  {\bibinfo {volume} {91}},\ \bibinfo {pages} {035136} (\bibinfo {year}
  {2015})}\BibitemShut {NoStop}%
\bibitem [{\citenamefont {Liu}\ \emph {et~al.}(2013{\natexlab{c}})\citenamefont
  {Liu}, \citenamefont {Repellin}, \citenamefont {Bernevig},\ and\
  \citenamefont {Regnault}}]{BeyondLaughlin}%
  \BibitemOpen
  \bibfield  {author} {\bibinfo {author} {\bibfnamefont {T.}~\bibnamefont
  {Liu}}, \bibinfo {author} {\bibfnamefont {C.}~\bibnamefont {Repellin}},
  \bibinfo {author} {\bibfnamefont {B.~A.}\ \bibnamefont {Bernevig}}, \ and\
  \bibinfo {author} {\bibfnamefont {N.}~\bibnamefont {Regnault}},\ }\href
  {\doibase 10.1103/PhysRevB.87.205136} {\bibfield  {journal} {\bibinfo
  {journal} {Phys. Rev. B}\ }\textbf {\bibinfo {volume} {87}},\ \bibinfo
  {pages} {205136} (\bibinfo {year} {2013}{\natexlab{c}})}\BibitemShut
  {NoStop}%
\bibitem [{\citenamefont {L\"auchli}\ \emph {et~al.}(2013)\citenamefont
  {L\"auchli}, \citenamefont {Liu}, \citenamefont {Bergholtz},\ and\
  \citenamefont {Moessner}}]{hierarchy}%
  \BibitemOpen
  \bibfield  {author} {\bibinfo {author} {\bibfnamefont {A.~M.}\ \bibnamefont
  {L\"auchli}}, \bibinfo {author} {\bibfnamefont {Z.}~\bibnamefont {Liu}},
  \bibinfo {author} {\bibfnamefont {E.~J.}\ \bibnamefont {Bergholtz}}, \ and\
  \bibinfo {author} {\bibfnamefont {R.}~\bibnamefont {Moessner}},\ }\href
  {\doibase 10.1103/PhysRevLett.111.126802} {\bibfield  {journal} {\bibinfo
  {journal} {Phys. Rev. Lett.}\ }\textbf {\bibinfo {volume} {111}},\ \bibinfo
  {pages} {126802} (\bibinfo {year} {2013})}\BibitemShut {NoStop}%
\bibitem [{\citenamefont {Wen}\ and\ \citenamefont {Niu}(1990)}]{wen}%
  \BibitemOpen
  \bibfield  {author} {\bibinfo {author} {\bibfnamefont {X.~G.}\ \bibnamefont
  {Wen}}\ and\ \bibinfo {author} {\bibfnamefont {Q.}~\bibnamefont {Niu}},\
  }\href {\doibase 10.1103/PhysRevB.41.9377} {\bibfield  {journal} {\bibinfo
  {journal} {Phys. Rev. B}\ }\textbf {\bibinfo {volume} {41}},\ \bibinfo
  {pages} {9377} (\bibinfo {year} {1990})}\BibitemShut {NoStop}%
\bibitem [{\citenamefont {Tao}\ and\ \citenamefont {Haldane}(1986)}]{quasideg}%
  \BibitemOpen
  \bibfield  {author} {\bibinfo {author} {\bibfnamefont {R.}~\bibnamefont
  {Tao}}\ and\ \bibinfo {author} {\bibfnamefont {F.~D.~M.}\ \bibnamefont
  {Haldane}},\ }\href {\doibase 10.1103/PhysRevB.33.3844} {\bibfield  {journal}
  {\bibinfo  {journal} {Phys. Rev. B}\ }\textbf {\bibinfo {volume} {33}},\
  \bibinfo {pages} {3844} (\bibinfo {year} {1986})}\BibitemShut {NoStop}%
\bibitem [{\citenamefont {Niu}\ \emph {et~al.}(1985)\citenamefont {Niu},
  \citenamefont {Thouless},\ and\ \citenamefont {Wu}}]{sflow1}%
  \BibitemOpen
  \bibfield  {author} {\bibinfo {author} {\bibfnamefont {Q.}~\bibnamefont
  {Niu}}, \bibinfo {author} {\bibfnamefont {D.~J.}\ \bibnamefont {Thouless}}, \
  and\ \bibinfo {author} {\bibfnamefont {Y.-S.}\ \bibnamefont {Wu}},\ }\href
  {\doibase 10.1103/PhysRevB.31.3372} {\bibfield  {journal} {\bibinfo
  {journal} {Phys. Rev. B}\ }\textbf {\bibinfo {volume} {31}},\ \bibinfo
  {pages} {3372} (\bibinfo {year} {1985})}\BibitemShut {NoStop}%
\bibitem [{\citenamefont {Laughlin}(1981)}]{LaughlinArgument}%
  \BibitemOpen
  \bibfield  {author} {\bibinfo {author} {\bibfnamefont {R.~B.}\ \bibnamefont
  {Laughlin}},\ }\href {\doibase 10.1103/PhysRevB.23.5632} {\bibfield
  {journal} {\bibinfo  {journal} {Phys. Rev. B}\ }\textbf {\bibinfo {volume}
  {23}},\ \bibinfo {pages} {5632} (\bibinfo {year} {1981})}\BibitemShut
  {NoStop}%
\bibitem [{\citenamefont {Haldane}(1985)}]{HaldaneCounting}%
  \BibitemOpen
  \bibfield  {author} {\bibinfo {author} {\bibfnamefont {F.~D.~M.}\
  \bibnamefont {Haldane}},\ }\href {\doibase 10.1103/PhysRevLett.55.2095}
  {\bibfield  {journal} {\bibinfo  {journal} {Phys. Rev. Lett.}\ }\textbf
  {\bibinfo {volume} {55}},\ \bibinfo {pages} {2095} (\bibinfo {year}
  {1985})}\BibitemShut {NoStop}%
\bibitem [{\citenamefont {Bernevig}\ and\ \citenamefont
  {Regnault}(2012)}]{BernevigCounting}%
  \BibitemOpen
  \bibfield  {author} {\bibinfo {author} {\bibfnamefont {B.~A.}\ \bibnamefont
  {Bernevig}}\ and\ \bibinfo {author} {\bibfnamefont {N.}~\bibnamefont
  {Regnault}},\ }\href {\doibase 10.1103/PhysRevB.85.075128} {\bibfield
  {journal} {\bibinfo  {journal} {Phys. Rev. B}\ }\textbf {\bibinfo {volume}
  {85}},\ \bibinfo {pages} {075128} (\bibinfo {year} {2012})}\BibitemShut
  {NoStop}%
\bibitem [{\citenamefont {Bergholtz}\ and\ \citenamefont
  {Liu}(2013)}]{RevBergholtz}%
  \BibitemOpen
  \bibfield  {author} {\bibinfo {author} {\bibfnamefont {E.~J.}\ \bibnamefont
  {Bergholtz}}\ and\ \bibinfo {author} {\bibfnamefont {Z.}~\bibnamefont
  {Liu}},\ }\href {\doibase 10.1142/S021797921330017X} {\bibfield  {journal}
  {\bibinfo  {journal} {International Journal of Modern Physics B}\ }\textbf
  {\bibinfo {volume} {27}},\ \bibinfo {pages} {1330017} (\bibinfo {year}
  {2013})}\BibitemShut {NoStop}%
\bibitem [{\citenamefont {Parameswaran}\ \emph {et~al.}(2013)\citenamefont
  {Parameswaran}, \citenamefont {Roy},\ and\ \citenamefont
  {Sondhi}}]{RevParameswaran}%
  \BibitemOpen
  \bibfield  {author} {\bibinfo {author} {\bibfnamefont {S.~A.}\ \bibnamefont
  {Parameswaran}}, \bibinfo {author} {\bibfnamefont {R.}~\bibnamefont {Roy}}, \
  and\ \bibinfo {author} {\bibfnamefont {S.~L.}\ \bibnamefont {Sondhi}},\
  }\href {\doibase http://dx.doi.org/10.1016/j.crhy.2013.04.003} {\bibfield
  {journal} {\bibinfo  {journal} {Comptes Rendus Physique}\ }\textbf {\bibinfo
  {volume} {14}},\ \bibinfo {pages} {816 } (\bibinfo {year}
  {2013})}\BibitemShut {NoStop}%
\bibitem [{\citenamefont {{Neupert}}\ \emph {et~al.}(2014)\citenamefont
  {{Neupert}}, \citenamefont {{Chamon}}, \citenamefont {{Iadecola}},
  \citenamefont {{Santos}},\ and\ \citenamefont {{Mudry}}}]{RevNeupert}%
  \BibitemOpen
  \bibfield  {author} {\bibinfo {author} {\bibfnamefont {T.}~\bibnamefont
  {{Neupert}}}, \bibinfo {author} {\bibfnamefont {C.}~\bibnamefont {{Chamon}}},
  \bibinfo {author} {\bibfnamefont {T.}~\bibnamefont {{Iadecola}}}, \bibinfo
  {author} {\bibfnamefont {L.~H.}\ \bibnamefont {{Santos}}}, \ and\ \bibinfo
  {author} {\bibfnamefont {C.}~\bibnamefont {{Mudry}}},\ }\href@noop {}
  {\bibfield  {journal} {\bibinfo  {journal} {ArXiv e-prints}\ } (\bibinfo
  {year} {2014})},\ \Eprint {http://arxiv.org/abs/1410.5828} {arXiv:1410.5828
  [cond-mat.str-el]} \BibitemShut {NoStop}%
\bibitem [{\citenamefont {Haque}\ \emph {et~al.}(2007)\citenamefont {Haque},
  \citenamefont {Zozulya},\ and\ \citenamefont {Schoutens}}]{Zozulya1}%
  \BibitemOpen
  \bibfield  {author} {\bibinfo {author} {\bibfnamefont {M.}~\bibnamefont
  {Haque}}, \bibinfo {author} {\bibfnamefont {O.}~\bibnamefont {Zozulya}}, \
  and\ \bibinfo {author} {\bibfnamefont {K.}~\bibnamefont {Schoutens}},\ }\href
  {\doibase 10.1103/PhysRevLett.98.060401} {\bibfield  {journal} {\bibinfo
  {journal} {Phys. Rev. Lett.}\ }\textbf {\bibinfo {volume} {98}},\ \bibinfo
  {pages} {060401} (\bibinfo {year} {2007})}\BibitemShut {NoStop}%
\bibitem [{\citenamefont {Zozulya}\ \emph {et~al.}(2007)\citenamefont
  {Zozulya}, \citenamefont {Haque}, \citenamefont {Schoutens},\ and\
  \citenamefont {Rezayi}}]{Zozulya2}%
  \BibitemOpen
  \bibfield  {author} {\bibinfo {author} {\bibfnamefont {O.~S.}\ \bibnamefont
  {Zozulya}}, \bibinfo {author} {\bibfnamefont {M.}~\bibnamefont {Haque}},
  \bibinfo {author} {\bibfnamefont {K.}~\bibnamefont {Schoutens}}, \ and\
  \bibinfo {author} {\bibfnamefont {E.~H.}\ \bibnamefont {Rezayi}},\ }\href
  {\doibase 10.1103/PhysRevB.76.125310} {\bibfield  {journal} {\bibinfo
  {journal} {Phys. Rev. B}\ }\textbf {\bibinfo {volume} {76}},\ \bibinfo
  {pages} {125310} (\bibinfo {year} {2007})}\BibitemShut {NoStop}%
\bibitem [{\citenamefont {Li}\ and\ \citenamefont
  {Haldane}(2008)}]{Entanglement}%
  \BibitemOpen
  \bibfield  {author} {\bibinfo {author} {\bibfnamefont {H.}~\bibnamefont
  {Li}}\ and\ \bibinfo {author} {\bibfnamefont {F.~D.~M.}\ \bibnamefont
  {Haldane}},\ }\href {\doibase 10.1103/PhysRevLett.101.010504} {\bibfield
  {journal} {\bibinfo  {journal} {Phys. Rev. Lett.}\ }\textbf {\bibinfo
  {volume} {101}},\ \bibinfo {pages} {010504} (\bibinfo {year}
  {2008})}\BibitemShut {NoStop}%
\bibitem [{\citenamefont {Peschel}(2003)}]{Peschel}%
  \BibitemOpen
  \bibfield  {author} {\bibinfo {author} {\bibfnamefont {I.}~\bibnamefont
  {Peschel}},\ }\href@noop {} {\bibfield  {journal} {\bibinfo  {journal}
  {Journal of Physics A: Mathematical and General}\ }\textbf {\bibinfo {volume}
  {36}},\ \bibinfo {pages} {L205} (\bibinfo {year} {2003})}\BibitemShut
  {NoStop}%
\bibitem [{\citenamefont {Turner}\ \emph {et~al.}(2010)\citenamefont {Turner},
  \citenamefont {Zhang},\ and\ \citenamefont {Vishwanath}}]{Vish:inv}%
  \BibitemOpen
  \bibfield  {author} {\bibinfo {author} {\bibfnamefont {A.~M.}\ \bibnamefont
  {Turner}}, \bibinfo {author} {\bibfnamefont {Y.}~\bibnamefont {Zhang}}, \
  and\ \bibinfo {author} {\bibfnamefont {A.}~\bibnamefont {Vishwanath}},\
  }\href {\doibase 10.1103/PhysRevB.82.241102} {\bibfield  {journal} {\bibinfo
  {journal} {Phys. Rev. B}\ }\textbf {\bibinfo {volume} {82}},\ \bibinfo
  {pages} {241102} (\bibinfo {year} {2010})}\BibitemShut {NoStop}%
\bibitem [{\citenamefont {Hermanns}\ \emph {et~al.}(2014)\citenamefont
  {Hermanns}, \citenamefont {Salimi}, \citenamefont {Haque},\ and\
  \citenamefont {Fritz}}]{Fritz:ES}%
  \BibitemOpen
  \bibfield  {author} {\bibinfo {author} {\bibfnamefont {M.}~\bibnamefont
  {Hermanns}}, \bibinfo {author} {\bibfnamefont {Y.}~\bibnamefont {Salimi}},
  \bibinfo {author} {\bibfnamefont {M.}~\bibnamefont {Haque}}, \ and\ \bibinfo
  {author} {\bibfnamefont {L.}~\bibnamefont {Fritz}},\ }\href
  {http://stacks.iop.org/1742-5468/2014/i=10/a=P10030} {\bibfield  {journal}
  {\bibinfo  {journal} {Journal of Statistical Mechanics: Theory and
  Experiment}\ }\textbf {\bibinfo {volume} {2014}},\ \bibinfo {pages} {P10030}
  (\bibinfo {year} {2014})}\BibitemShut {NoStop}%
\bibitem [{\citenamefont {Prodan}\ \emph {et~al.}(2010)\citenamefont {Prodan},
  \citenamefont {Hughes},\ and\ \citenamefont {Bernevig}}]{Bernevig:Disord}%
  \BibitemOpen
  \bibfield  {author} {\bibinfo {author} {\bibfnamefont {E.}~\bibnamefont
  {Prodan}}, \bibinfo {author} {\bibfnamefont {T.~L.}\ \bibnamefont {Hughes}},
  \ and\ \bibinfo {author} {\bibfnamefont {B.~A.}\ \bibnamefont {Bernevig}},\
  }\href {\doibase 10.1103/PhysRevLett.105.115501} {\bibfield  {journal}
  {\bibinfo  {journal} {Phys. Rev. Lett.}\ }\textbf {\bibinfo {volume} {105}},\
  \bibinfo {pages} {115501} (\bibinfo {year} {2010})}\BibitemShut {NoStop}%
\bibitem [{\citenamefont {Alexandradinata}\ \emph {et~al.}(2011)\citenamefont
  {Alexandradinata}, \citenamefont {Hughes},\ and\ \citenamefont
  {Bernevig}}]{Alex:CM}%
  \BibitemOpen
  \bibfield  {author} {\bibinfo {author} {\bibfnamefont {A.}~\bibnamefont
  {Alexandradinata}}, \bibinfo {author} {\bibfnamefont {T.~L.}\ \bibnamefont
  {Hughes}}, \ and\ \bibinfo {author} {\bibfnamefont {B.~A.}\ \bibnamefont
  {Bernevig}},\ }\href {\doibase 10.1103/PhysRevB.84.195103} {\bibfield
  {journal} {\bibinfo  {journal} {Phys. Rev. B}\ }\textbf {\bibinfo {volume}
  {84}},\ \bibinfo {pages} {195103} (\bibinfo {year} {2011})}\BibitemShut
  {NoStop}%
\bibitem [{\citenamefont {Kargarian}\ and\ \citenamefont
  {Fiete}(2010)}]{PhysRevB.82.085106}%
  \BibitemOpen
  \bibfield  {author} {\bibinfo {author} {\bibfnamefont {M.}~\bibnamefont
  {Kargarian}}\ and\ \bibinfo {author} {\bibfnamefont {G.~A.}\ \bibnamefont
  {Fiete}},\ }\href {\doibase 10.1103/PhysRevB.82.085106} {\bibfield  {journal}
  {\bibinfo  {journal} {Phys. Rev. B}\ }\textbf {\bibinfo {volume} {82}},\
  \bibinfo {pages} {085106} (\bibinfo {year} {2010})}\BibitemShut {NoStop}%
\bibitem [{\citenamefont {Sterdyniak}\ \emph {et~al.}(2011)\citenamefont
  {Sterdyniak}, \citenamefont {Regnault},\ and\ \citenamefont
  {Bernevig}}]{EntanglementExcitation}%
  \BibitemOpen
  \bibfield  {author} {\bibinfo {author} {\bibfnamefont {A.}~\bibnamefont
  {Sterdyniak}}, \bibinfo {author} {\bibfnamefont {N.}~\bibnamefont
  {Regnault}}, \ and\ \bibinfo {author} {\bibfnamefont {B.~A.}\ \bibnamefont
  {Bernevig}},\ }\href {\doibase 10.1103/PhysRevLett.106.100405} {\bibfield
  {journal} {\bibinfo  {journal} {Phys. Rev. Lett.}\ }\textbf {\bibinfo
  {volume} {106}},\ \bibinfo {pages} {100405} (\bibinfo {year}
  {2011})}\BibitemShut {NoStop}%
\bibitem [{\citenamefont {Sterdyniak}\ \emph {et~al.}(2012)\citenamefont
  {Sterdyniak}, \citenamefont {Regnault},\ and\ \citenamefont
  {M\"oller}}]{ParticleEntanglement}%
  \BibitemOpen
  \bibfield  {author} {\bibinfo {author} {\bibfnamefont {A.}~\bibnamefont
  {Sterdyniak}}, \bibinfo {author} {\bibfnamefont {N.}~\bibnamefont
  {Regnault}}, \ and\ \bibinfo {author} {\bibfnamefont {G.}~\bibnamefont
  {M\"oller}},\ }\href {\doibase 10.1103/PhysRevB.86.165314} {\bibfield
  {journal} {\bibinfo  {journal} {Phys. Rev. B}\ }\textbf {\bibinfo {volume}
  {86}},\ \bibinfo {pages} {165314} (\bibinfo {year} {2012})}\BibitemShut
  {NoStop}%
\bibitem [{\citenamefont {Weeks}\ and\ \citenamefont {Franz}(2010)}]{Weeks}%
  \BibitemOpen
  \bibfield  {author} {\bibinfo {author} {\bibfnamefont {C.}~\bibnamefont
  {Weeks}}\ and\ \bibinfo {author} {\bibfnamefont {M.}~\bibnamefont {Franz}},\
  }\href {\doibase 10.1103/PhysRevB.82.085310} {\bibfield  {journal} {\bibinfo
  {journal} {Phys. Rev. B}\ }\textbf {\bibinfo {volume} {82}},\ \bibinfo
  {pages} {085310} (\bibinfo {year} {2010})}\BibitemShut {NoStop}%
\bibitem [{\citenamefont {Ni\ifmmode \mbox{\c{t}}\else
  \c{t}\fi{}\ifmmode~\u{a}\else \u{a}\fi{}}\ \emph {et~al.}(2013)\citenamefont
  {Ni\ifmmode \mbox{\c{t}}\else \c{t}\fi{}\ifmmode~\u{a}\else \u{a}\fi{}},
  \citenamefont {Ostahie},\ and\ \citenamefont {Aldea}}]{Aldea}%
  \BibitemOpen
  \bibfield  {author} {\bibinfo {author} {\bibfnamefont {M.}~\bibnamefont
  {Ni\ifmmode \mbox{\c{t}}\else \c{t}\fi{}\ifmmode~\u{a}\else \u{a}\fi{}}},
  \bibinfo {author} {\bibfnamefont {B.}~\bibnamefont {Ostahie}}, \ and\
  \bibinfo {author} {\bibfnamefont {A.}~\bibnamefont {Aldea}},\ }\href
  {\doibase 10.1103/PhysRevB.87.125428} {\bibfield  {journal} {\bibinfo
  {journal} {Phys. Rev. B}\ }\textbf {\bibinfo {volume} {87}},\ \bibinfo
  {pages} {125428} (\bibinfo {year} {2013})}\BibitemShut {NoStop}%
\bibitem [{\citenamefont {Zhao}\ and\ \citenamefont {Shen}(2012)}]{Zhao}%
  \BibitemOpen
  \bibfield  {author} {\bibinfo {author} {\bibfnamefont {A.}~\bibnamefont
  {Zhao}}\ and\ \bibinfo {author} {\bibfnamefont {S.-Q.}\ \bibnamefont
  {Shen}},\ }\href {\doibase 10.1103/PhysRevB.85.085209} {\bibfield  {journal}
  {\bibinfo  {journal} {Phys. Rev. B}\ }\textbf {\bibinfo {volume} {85}},\
  \bibinfo {pages} {085209} (\bibinfo {year} {2012})}\BibitemShut {NoStop}%
\bibitem [{\citenamefont {Goldman}\ \emph {et~al.}(2011)\citenamefont
  {Goldman}, \citenamefont {Urban},\ and\ \citenamefont {Bercioux}}]{Goldman}%
  \BibitemOpen
  \bibfield  {author} {\bibinfo {author} {\bibfnamefont {N.}~\bibnamefont
  {Goldman}}, \bibinfo {author} {\bibfnamefont {D.~F.}\ \bibnamefont {Urban}},
  \ and\ \bibinfo {author} {\bibfnamefont {D.}~\bibnamefont {Bercioux}},\
  }\href {\doibase 10.1103/PhysRevA.83.063601} {\bibfield  {journal} {\bibinfo
  {journal} {Phys. Rev. A}\ }\textbf {\bibinfo {volume} {83}},\ \bibinfo
  {pages} {063601} (\bibinfo {year} {2011})}\BibitemShut {NoStop}%
\bibitem [{\citenamefont {Parameswaran}\ \emph {et~al.}(2012)\citenamefont
  {Parameswaran}, \citenamefont {Roy},\ and\ \citenamefont
  {Sondhi}}]{BerryParameswaran}%
  \BibitemOpen
  \bibfield  {author} {\bibinfo {author} {\bibfnamefont {S.~A.}\ \bibnamefont
  {Parameswaran}}, \bibinfo {author} {\bibfnamefont {R.}~\bibnamefont {Roy}}, \
  and\ \bibinfo {author} {\bibfnamefont {S.~L.}\ \bibnamefont {Sondhi}},\
  }\href {\doibase 10.1103/PhysRevB.85.241308} {\bibfield  {journal} {\bibinfo
  {journal} {Phys. Rev. B}\ }\textbf {\bibinfo {volume} {85}},\ \bibinfo
  {pages} {241308} (\bibinfo {year} {2012})}\BibitemShut {NoStop}%
\bibitem [{\citenamefont {Kourtis}\ \emph {et~al.}(2014)\citenamefont
  {Kourtis}, \citenamefont {Neupert}, \citenamefont {Chamon},\ and\
  \citenamefont {Mudry}}]{FarExceed}%
  \BibitemOpen
  \bibfield  {author} {\bibinfo {author} {\bibfnamefont {S.}~\bibnamefont
  {Kourtis}}, \bibinfo {author} {\bibfnamefont {T.}~\bibnamefont {Neupert}},
  \bibinfo {author} {\bibfnamefont {C.}~\bibnamefont {Chamon}}, \ and\ \bibinfo
  {author} {\bibfnamefont {C.}~\bibnamefont {Mudry}},\ }\href {\doibase
  10.1103/PhysRevLett.112.126806} {\bibfield  {journal} {\bibinfo  {journal}
  {Phys. Rev. Lett.}\ }\textbf {\bibinfo {volume} {112}},\ \bibinfo {pages}
  {126806} (\bibinfo {year} {2014})}\BibitemShut {NoStop}%
\bibitem [{\citenamefont {Grushin}\ \emph {et~al.}(2012)\citenamefont
  {Grushin}, \citenamefont {Neupert}, \citenamefont {Chamon},\ and\
  \citenamefont {Mudry}}]{SingleParticleGrushin}%
  \BibitemOpen
  \bibfield  {author} {\bibinfo {author} {\bibfnamefont {A.~G.}\ \bibnamefont
  {Grushin}}, \bibinfo {author} {\bibfnamefont {T.}~\bibnamefont {Neupert}},
  \bibinfo {author} {\bibfnamefont {C.}~\bibnamefont {Chamon}}, \ and\ \bibinfo
  {author} {\bibfnamefont {C.}~\bibnamefont {Mudry}},\ }\href {\doibase
  10.1103/PhysRevB.86.205125} {\bibfield  {journal} {\bibinfo  {journal} {Phys.
  Rev. B}\ }\textbf {\bibinfo {volume} {86}},\ \bibinfo {pages} {205125}
  (\bibinfo {year} {2012})}\BibitemShut {NoStop}%
\end{thebibliography}%
\end{document}